\documentclass[a4paper,11pt,final]{article}
\usepackage{graphicx} 
\usepackage[applemac]{inputenc}

\author{B. Bassetti\footnote{Università Statale di Milano, Via Celoria 
16, 20100 Milano, Italy; e-mail address bassetti@mi.infn.it}, M. 
Cosentino Lagomarsino\footnote{Università Statale di Milano (Will
take the address: FOM Institute for Atomic and Molecular Physics
(Amolf), Kruislaan 407, 1098 SJ Amsterdam, The Netherlands}, P. Jona 
\footnote{Politecnico di Milano, Pza Leonardo Da Vinci 1, 20100 
Milano, Italy}}

\title{A Model for the Self-Organization of 
Microtubules Driven by Molecular Motors}

\begin{document}
   
\maketitle

{\large 
\begin{center}
\end{center}
}

\begin{abstract}
    We propose a two-dimensional model for the organization of
    stabilized microtubules driven by molecular motors in an
    unconfined geometry.  In this model two kinds of dynamics are
    competing.  The first one is purely diffusive, with an interaction
    between the rotational degrees of freedom, while the second one is
    a local drive, dependent on microtubule polarity.  As a result,
    there is a configuration dependent driving field.  Applying a
    molecular field approximation, we are able to derive continuum
    equations.  A study on the solutions of these equations shows
    nonequilibrium inhomogeneous steady states in various regions of
    the parameter space.  The presence and stability of such
    self-organized states are investigated in terms of entropy
    production.  Numerical simulations confirm our analytic results.
\end{abstract}

\section{Introduction}\label{sec.1}
Using a theoretical physics approach, we introduce and analyze a 
stochastic model inspired by the self assembly process of the 
cytoskeleton. 

The cytoskeleton can be seen as the ``infrastructure'' of eukaryotic
cells, providing for both (dynamically evolving) spatial structure and
internal transport processes that are fundamental for the cell itself
and its role in a living organism (see~\cite{watson}).  We focus on
questions concerning the statistical mechanics of this particular
biological system, mainly its ability to assemble in a variety of
symmetry-breaking phases, which are commonly believed to have
nonequilibrium nature and are considered important for cell
morphogenesis.  The nonequilibrium forces that give rise to these
phases are usually ascribed to both the so called \emph{dynamic
instability} of microtubules (associated with a confining geometry,
see for example \cite{marileen}) and the action of molecular
motors.  Dynamic instability is a nonequilibrium
polimerization/depolimerization process that enables microtubules or
actin filaments to exert forces on confining surfaces such as the cell
membrane.  Molecular motors are a much studied family of proteins that
are able to generate active motion on cytoskeletal filaments. 

In vitro and numerical experiments, sometimes
called self-organization assays (\cite{heald}, \cite{ned1}, 
\cite{ned2}, \cite{gibbons}), show that ensembles of microtubules and 
active motors can self-organize under proper conditions, and this ability 
is strictly linked to motor activity. In these assays, motors are 
typically found in soluble complexes and operate when two or more 
filaments are present; dynamic instability doesn't seem to be determinant. 

In this paper we concentrate on the role of motor proteins and do not 
consider dynamic instability.
In a previous article (\cite{faretta}) we have shown
theoretically that in motility assays, where motor proteins are
adsorbed on surfaces and the ``gliding'' process of single
microtubules is observed, rotational diffusion \cite{dire} leads to
purely diffusive dynamics.

This fact brought us to investigate the breaking of rotational
symmetry, in the form of pattern formation, as a many body
effect. That is, we consider the excluded volume
interaction of many filaments. This is different 
from cooperation of motor proteins that
is typically investigated in the situation of muscle contraction; (see
\cite{prost} and \cite{frey} for interesting recent examples of
stochastic modeling in this context). 
 
Concisely, our problem can be stated in terms 
of existence of inhomogeneous nonequilibrium steady states.
From this viewpoint, our model is on the same level as many other 
models of non-biological systems, such as a fluid with convecting flow, or a 
nonequilibrium chemical reactor (this is true as long as we 
are not modeling the self-regulatory processes that take part in cell 
morphogenesis, see \cite{morfo}).

What makes our choice distinct and particularly interesting for
statistical mechanics is that the generalized force which keeps the
system far from equilibrium is not, in general, a global field or some
boundary condition, as is usually found in the literature on far from
equilibrium systems, but a local release of energy associated with
transport.

The model, which is a two-dimensional lattice spin system, is 
presented in section \ref{sec.2}.  

In order to discover the relevant variables of the real system we
explored different ways to implement its microscopic dynamics \cite{marco}.  
In this paper we limit the discussion to the simplest case we could 
find showing significant results. 

This point needs some clarification.  Our working hypothesis is
that the relevant features of the system are the competition between 
diffusion and motor drive, together with excluded volume interactions.
The spirit of our investigation is not to reproduce in
detail the mechanical features of the microscopic system but to
analyze the behavior of averaged quantities. Many different
dynamics can reproduce the right averages (we follow the
concept of universality in statistical mechanics), so our job is to
find the essential features that a microscopic dynamics needs to have to
reproduce the desired behavior. Of course this leaves
many possibilities open to choose a microscopic dynamics and one has
to be guided by considerations on experimental models.  These
considerations are extremely useful to give an intuitive grasp on the
interpretation of microscopic dynamics.  Nevertheless, the rigorous
justification of our microscopic model has to be sought in the results
it gives for macroscopic, averaged observables.

All of the dynamics we developed are based on the 
competition between a diffusive process (when motors are detached from 
filaments or not active) and a driven-diffusive one (when motors are 
active), this means that they fall in the class of competing dynamics 
models (\cite{mendes}, \cite{garrido}).

In the system presented in this paper, the driven 
process is realized in a way that is resemblant to B. Schmittmann and 
R.K.P. Zia's \emph{driven diffusive systems} (\cite{zia1}, 
\cite{zia2}). This case is most easily interpreted in terms of many 
body motility assays, so we will stick to this interpretation 
throughout the paper. Unfortunately, we are not aware of any 
experiments of this kind focused on self-organization, so we dedicate section 
\ref{sec.6} to discuss the possibilities of employing this microscopic 
dynamics for the experiment in reference \cite{ned1} and the results 
we obtained with a different choice for the driven process, which 
seemes more sound for this case.    

In section \ref{sec.3}, starting from microscopic dynamics, we use a 
mean field approximation to derive a set of four discrete evolution 
equations for the system.  We discuss the problems that arise when one 
considers the continuum limit of these equations in order to obtain a 
more at hand system of differential equations.  In particular, the 
result of this limit is related to the characteristic times of the 
two competing dynamics.

Once we have obtained a set of differential equations on $ 
\mathcal{R}^{2} $, we look for steady states in different regimes.  
In the two limiting cases of dominant diffusion and dominant drive we 
find, respectively, the usual homogeneous Gibbs states and blocked 
phases; whereas, when the two dynamics are mixed, a new class of 
entropy-producing, phase-separated states emerges.  By linear 
stability analysis, we find in parameter space the instability region 
of the homogeneous states.

Finally, in section \ref{sec.5}, these results are compared with those 
obtained by numerical simulation. In order to tackle the problem of 
entropy production numerically, we apply some of the machinery 
related to the \emph{fluctuation theorem} by Gallavotti and Cohen 
(reviewed in \cite{galla}), that has recently been established for 
stochastic dynamics (\cite{kurchan}, \cite{lebo}, \cite{maes}).

\section{Microscopic Dynamics}\label{sec.2}

We imagine that motors are adsorbed on a flat surface and can push the
filaments as in motility assays.  Our aim is to find a transition to
self-organized, inhomogeneous states.  Experiments of this kind,
though in principle possible, have not been tried to our knowledge.

On the other hand, in the self-organization experiments described 
in \cite{ned1} and \cite{ned2}, microtubule-associated motors are 
linked in bundles by some other proteins, and these bundles are in
solution. Links of these experiments to our model will be discussed in
section \ref{sec.6}.

\begin{figure}[!hb]
    \centering
    \includegraphics{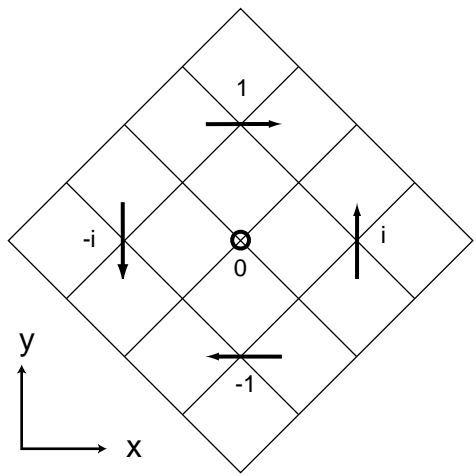}
	\caption{Conventions adopted for lattice and occupation variables.  
	Complex values of the spins are merely adopted to make our 
	computational life easier.  Microtubule directions are along the axes.  
	Spins are \protect $ 45 \circ \protect $ to lattice bonds so 
	that interaction 
	with all four nearest neighbors is symmetric.}
    \label{fig.1}
\end{figure}

We discretize space, time, and the orientation of the filaments. 
So we define a 2-dimensional square lattice $ \Lambda
$ and we imagine that each lattice site is either empty or occupied by
the center of mass of a filament.
This corresponds to associate to each lattice site $ x \in \Lambda $ 
a spin $ \sigma_{x}$ which takes the null value when the site is
empty. The other possible values of the spins are determined by the
discretization on the orientational degree of freedom.  With these
assumptions, microtubules are treated as rigid rods, neglecting 
dynamic instability. The minimal choice is that spins take values in
$\{+1,-1,+\mathrm{i},-\mathrm{i},0\} $, 
corresponding to the four fundamental orientations of the filaments and to
the empty site (see fig. \ref{fig.1}). Our results will show that
this choice is significant.
    
The thermodynamic equilibrium properties of the system are 
determined entirely by its hamiltonian $ H $. We take

\begin{equation}
   H = J\sum_{\stackrel{x,y}{\mathrm{ n.n.}}} \sigma^{2}_{x} 
   \sigma^{2}_{y} + K\sum_{\stackrel{x,y}{\mathrm{ n.n.}}} 
   \sigma^{4}_{x} \sigma^{4}_{y}
        \label{eq:1}
\end{equation}

This is just the most generic form that $ H $ can take in our case,
provided that the interaction is nearest neighbor.  

Intuitively, the first coupling constant, $ J < 0$, stands for an interaction
between the directions of the rods (insensitive to their
orientations), and mimics excluded volume effects on directions.  $ K
$, on the other hand, is sensitive to the presence or absence of the 
filaments.

The choice of a nearest neighbor hamiltonian is connected to the
problem of the length of the filaments. Reasoning geometrically, if $
a $ is the lattice spacing, a filament of length $ L $ should interact with $
\frac{L}{a} $ sites. On the other hand, we can do the
statistical mechanics of this system after a preliminary coarse
graining procedure. This renders the hamiltonian nearest neighbors,
while the constants  \label{lunghezza} $ J $ and $ K $ keep memory
of the original scale, becoming functions of the (mean) length $ L $
of the filaments (see for example \cite{degennes}). The spin of each
site takes the meaning of a cluster spin, so even though we speak of 
filaments as they were in $ 1:1 $ correspondence with our spins, 
this is merely a convention, because one spin may already be the 
result of mesoscopic averaging procedures.

From expression (\ref{eq:1}) it is easy to see that the statics of
this  model is equivalent to that of the 2-D Blume-Emery-Griffiths
model with only  two relevant coupling constants (see \cite{beg} for a
complete  mean-field analysis of the phase diagram).  In fact, the
effective  levels reduce to three ($ \sigma^{2} = \pm 1, 0 $), two
of which are  2-fold degenerate.  They correspond to the two possible
directions  (regardless of orientation) of the filaments and their
absence.

Since external fields are absent, our hamiltonian produces two  kinds
of spatially isotropic states, which are associated respectively  with
the well-known liquid/gas and isotropic/nematic transitions
(\cite{beg}). Notice that the phases that the system can exhibit are
related with the assumption on the discretization of the rotational
degrees of freedom. A finer discretization could bring to a larger
number of phases. Anyway, the latter would be all be spatially
isotropic, causing little change in the problem of the transition to
inhomogeneous states.

We now consider the time-dependent properties of the model.  The
evolution algorithm is made of two branches, of which the first is a
diffusion and the second a driven diffusive process. Both processes
take place in presence of the hamiltonian $ H $ and include two
kinds of elementary Montecarlo moves:

\begin{itemize}
    \item Nearest neighbor spin exchange (only effective when one of
    the two sites is empty) $ \sigma_{x} \: \leftrightarrow
    \sigma_{x'} \quad x' \in \mathrm{n.n.}(x)$ (Kawasaki dynamics).

    \item Local orientation $ \sigma_{x} \ne 0 \rightarrow
    \sigma_{x}' \ne 0$ (Glauber dynamics).
\end{itemize}

In the diffusive steps, when the interaction with motors is not
active, these two kinds of moves correspond respectively to
translational and rotational diffusion of a microtubule.

The probability of accepting a move is a modified Metropolis \[
\mathcal{A} = \frac{1}{2}(1 + \tanh \frac{\Delta H}{2})\]

Let's now ``turn on'' the interaction with motor proteins, and see
what happens.  What we assume is that motor drive will act on the
translational diffusion of the centers of mass of the filaments,
trasforming it in a driven diffusion, and that rotational diffusion
will be inhibited.

\begin{figure}[!htbp]
    \centering 
	\includegraphics{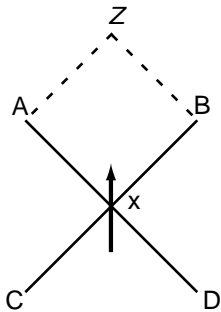} 
		\caption{Motor action. 
	\protect$ \Delta H \protect$ is shifted by \protect$ -E \protect$ for  
Kawasaki exchanges with sites \protect$ A \protect$ and \protect$ B \protect$, 
identified by  the arrow head, and by
 \protect$ +E \protect$ for exchanges with \protect$ C \protect$
	and 
\protect$  D \protect$.}  \label{fig:2} \end{figure}

In order to model motor activity, we modify $ \mathcal{A} $ in the
same way that is commonly used in the context of driven diffusive
systems \cite{zia1}.  That is, $ \Delta H(C,C') \rightarrow \Delta
H(C,C') \pm E(C)$ when configuration $ C $ is favorable for motor
pushing forward or pulling back the filament (see figure \ref{fig:2}).
$ H $ can be taken to be the same as in (\ref{eq:1}). $ E(C) $ is
the driving field, and is proportional to the work done by the motors.

This way of modeling molecular motor driven dynamics contains a number
of hypotheses.  First of all, we suppose that the motors will push the
microtubules every time they can. Secondly, the motors are spread with
constant density and push one filament at a time (this last choice is
intuitive for motility assays; in section \ref{sec.6} it will be
discussed for the  case of self-organization assays).  Lastly, the
motor action constrains the microtubules to preserve their
orientations.  The justification for this last hypothesis for a
motility assay lies in the fact that as soon as two motors are
attached to one filament, its direction is frozen except for the
elastic fluctuations of its head and tail (see \cite{leibler} and
\cite{faretta}).

The fact that the pushing of the motor is actually a diagonal
translation may cause some perplexities. On the other hand, our
dynamics is set up taking into account the fact that we have two
degrees of freedom, translational and rotational, that are
independent. So they have to be activated in different elementary
moves. The overall motion springs from a sequence of these moves. In
this wiew, motor action is just a bias in the translational dynamics.
We also tested (\cite{marco}) numerically a dynamics where motors push linearly
(exchange is favoured with site $ Z $ in figure \ref{fig:2})
obtaining no qualitative change in the results (but this choice
complicates our analytic mean field calculations because of the fact
that translations involve next nearest neighbours). \label{varia1}

We would like to make a final remark on the boundary conditions.  We
adopted periodic boundary conditions in our simulations.

The simultaneous presence of two competing dynamics, each one with its
own characteristic time, makes it reasonable to expect (see
\cite{garrido}) that possible steady states have a nonequilibrium (non
Gibbsian) nature independently on the boundary conditions.  This fact
is confirmed by our mean field analysis.
  
When considering the dynamics of active motors alone, the condition
above ceases to be valid and we are in the case of a four species
driven  system. One should keep in mind the following remark.

When $ E \ne 0 $, the resulting driving field depends on the
spatiotemporal configuration of the system.  As it arises from a
modification of the Metropolis acceptation probability, each
translational step of the microscopic dynamics automatically satisfies
detailed balance with respect to the modified energy difference,

\begin{displaymath}
	\frac{W(C'\to C)}{W(C\to C')} = \frac{P_{eq}(C)}{P_{eq}(C')} =
	e^{\beta (\Delta H \pm E(C) )}
\end{displaymath}


However, in general, detailed balance is not satisfied for periodic
boundary conditions \cite{zia1}.  For this reason, one can expect to
observe inhomogeneous (``blocked'') steady states even in the presence
of the sole motor dynamics.

To sum up, these are the main features of the model:

It is a model with competing dynamics.  Thus, the characteristic time
scale ratio of the two dynamics plays  an important role in its
behavior \cite{mendes}.

It is related to so-called multiple species driven systems
(\cite{zia1}, \cite{zia2}, \cite{foster}), to which it reduces when
Glauber-like dynamics (rotational diffusion) is not active and
periodic boundary conditions are set.

It can be interpreted as a reaction/diffusion model, in which  four
diffusing species, driven in four orthogonal directions, are  subject
to (equilibrium) chemical reactions controlled by the Ising
hamiltonian (\ref{eq:1}).

\section{Mean field approximation and continuum equations}\label{sec.3}

The dynamics we described in the preceding section defines a Markov
process on the lattice $ \Lambda $.  The time-dependent probability
of a configuration $ C = \{ \sigma_{x} \}_{x \in \Lambda}$ is given
by the master equation (\cite{prigo}, \cite{garrido})

\begin{equation}
    	\frac{d}{d t} P(C,t) = (\tau_{p}\mathbf{L}_{p} +
    	\tau_{a}\mathbf{L}_{a}) P(C,t) \label{eq:2}
\end{equation}

where $ \tau_{p} $, $ \tau_{a} $ are the characteristic times of
the two dynamics.  Physically, $\tau_{a}$ and $\tau{p}$ represent
a slow process with respect to the entire process of interaction
between a motor and a filament.  Therefore at the time scale we adopt,
the dynamics is regarded as diffusive or driven respectively when a
filament encounters on average a small or a large number of motors.

$ \mathbf{L}_{p} $, $ \mathbf{L}_{a} $ are the generators of the
evolution when the motors are passive and active respectively.  These
operators contain the transition probabilities $ W(C,C') $ from
configuration $ C $ to $ C' $, which can be easily written down
following the description of the model given in the preceding section.
For diffusive dynamics regulated by hamiltonian (\ref{eq:1}) (switched
off motors)

\begin{displaymath}
    \begin{array}{cc} W(x| C\to C') \quad = \qquad \qquad \quad & \\
		=
		\quad
		\delta_{\sigma_{x},0}\delta_{\sigma_{x},\sigma'_{x}}
		+  \frac{1}{4}(1 - \delta_{\sigma_{x},0}) \quad \times
		&\\ \times \sum_{q}\{ (1 -  \delta_{\sigma_{x+q},0})
		\frac{1}{3} (1 -  \delta_{\sigma'_{x},0})(1 -
		\delta_{\sigma'_{x},\sigma_{x}}) & \! [  \frac{1}{2} (
		1 - \mathrm{Th} \frac{\Delta H}{2})] +  \\ + \; \;  (
		\delta_{\sigma_{x+q},0}) \delta_{\sigma'_{x},0}
		\delta_{\sigma'_{x+q},\sigma_{x}} [  \frac{1}{2} ( 1 -
		\mathrm{Th} \frac{\Delta H}{2})] \} & \end{array}
\end{displaymath}

In the case of driven dynamics the corrections are that Glauber
transitions (rotations) are inhibited, and $ \Delta H(C,C')
\rightarrow \Delta  H(C,C') \pm E(C)$ when the configuration is such
that the motors can do  work (fig.  \ref{fig:2}).

Instead of postulating, as is usually done \cite{zia1}, some
mesoscopic equations on the basis of symmetries and physical
considerations, we start from microscopic dynamics and develop a
local mean field approximation. The main assertion of this
approximation is that the probability of a configuration factorizes as

\begin{displaymath}
    P({\sigma_{x}}) = \prod_{x \in \Lambda} p(\sigma_{x})
\end{displaymath}

where $ p(\sigma_{x}) $ is the most general single site measure:
\begin{displaymath}
    p(\sigma_{x}) = \sum_{I} p_{I}\delta_{\sigma_{x},I} \quad ; \quad
    \sum_{I}p_{I} = 1
\end{displaymath}
($ I $ runs on all the possible values of the spin).

With these assumptions, the mean value of a function of $ n $ spins
can be written
\begin{displaymath}
    \langle F(\sigma_{x_{1}},\ldots,\sigma_{x_{n}}) \rangle =
    \sum_{I_{1},\ldots,I_{n}}F(I_{1},\ldots,I_{n})[p_{I_{1}}(x_{1})\ldots
    p_{I_{n}}(x_{n})]
\end{displaymath}

We use this approximation to obtain the evolution equations of the
first moments.  Such discrete equations are sums of the mean values of
some quantities computed in four different sites of the lattice,
reaching next-next nearest neighbors.  Namely, given a lattice versor
$ \mathbf{a} $, one has to consider the cluster containing both the
nearest neighbors of $ x $ and those of $ x + \mathbf{a} $.  We
get expressions that involve the parameters

\begin{itemize}
	\item $ H(x) = \langle  1 - \sigma_{x}^{4} \rangle$
        (density of holes).

	\item $ M(x)  = \langle \sigma_{x}^{2} \rangle$ (quadrupole
	moment;  interpreted as the mean value of the net filament
	direction).

	\item $ g(x) + \mathrm{i} f(x) = \langle \sigma_{x}
	\rangle$, identified with vector field $ \vec{D}(x) = (g(x),
	f(x)) $ (orientation of the filaments).
\end{itemize}

Using Taylor expansions we obtain the following equations:

{\linespread{1.8}
\begin{equation}
	     {\displaystyle \begin{array}{c}  \dot{H} = - \:
		 \mathrm{div}[ -  \nabla H + J H M\nabla M - K H
		 (1-H)\nabla H] \: \: + \\ - \: \: r E \mathrm{div} (H
		 \vec{D}) \qquad \qquad \qquad  \qquad \qquad \quad
		 \end{array}} \label{eq:3}
\end{equation}
\begin{equation}
        {\displaystyle \begin{array}{c}  \dot{M} =  \: \mathrm{div}[
        (H)^{2}\nabla\frac{M}{H} +  J H (1-H)\nabla M - K H M \nabla H
        ] \: \: + \\ + \: \: r E \mathrm{div}(H \hat{T} \vec{D})] \:
        \: + \qquad  \qquad \qquad \qquad \: \\ - \: \: (1 - r) (M +
        \mathrm{Th}(4JM) ) \qquad  \qquad\: \: \end{array}}
        \label{eq:4}
\end{equation}
\begin{equation}
        {\displaystyle \begin{array}{c} \dot{g} = \: \: \mathrm{div}[
            H^{2}\nabla  \frac{g}{H}   + J g H \nabla M  - K g H
            \nabla H ] \: \: + \\ + \: \: r E \partial_{x}[(1-H+M)H]
            \: \: + \qquad \quad \\ - \: \:  (1 -r) g (1 +
            \frac{1}{2}\mathrm{Th}(4JM) )  \qquad \end{array}}
            \label{eq:5}
\end{equation}
\begin{equation}
    {\displaystyle     \begin{array}{c} \dot{f} = \: \: \mathrm{div}[
             H^{2}\nabla  \frac{f}{H}   - J f H \nabla M  - K f N
             \nabla H ] \: \: + \\ - \: \: r E \partial_{y}[(1-H-M)H]
             \: \: + \qquad \quad \\ - \: \:  (1-r) f (1 -
             \frac{1}{2}\mathrm{Th}(4JM) ) \qquad \end{array}}
             \label{eq:6}
\end{equation}
} Where $ \hat{T} $ is the conjugate operator in $ R^2 $.

A continuum limit has been performed along an appropriate path in
parameter space, so that one characteristic time and the lattice
spacing $ a $ have been eliminated.  Consequently, we are left with
four significant (rescaled) parameters, $ J $, $ K $, $ E $, $
r $.  The first two concern the equilibrium properties of the
system, while the others are indicators of its being far from
equilibrium.

$ r $ gives the relative weight of the two dynamics and is related
to the ratio $ \Gamma =\frac{\tau_{p}}{\tau_{a}}$ of the two
characteristic times by the equality $ r = \frac{1}{\Gamma + 1} $.
$r$ is independent from the motor drive $E$; as matter of fact,
$E$ only appears in association with $r$ in the equations, so that
we interpret the quantity $ r E $ as the nonequilibrium generalized
force generated by the motor action.

In the equations above, $ r $ may be regarded as a free,
phenomenological, parameter, as is not fixed by any assumption on our
model.

Establishing the relationship between $ r $ and some underlying
microscopic parameters, such as motor density and microtubule length,
is  a problem that falls beyond the reach of our model, and that
requires  a more detailed analysis of the physical system.

For motility assays, the analysis of T.Duke and others
(\cite{leibler}) specifies the microscopic process as a Markovian
stochastic evolution of the number $ n $ of motor proteins attached
to a  microtubule. Following this work, one can identify $ \tau_{a}
$  with the mean first passage time for going from $ n = 2 $ (two
attached motors) to $ n = 1 $ (one attached motor), and $ \tau_{p}
$  with the mean time in which a microtubule is attached to $ 0 $
or $  1 $ motor, obtaining:

\begin{displaymath}
  \tau_{a} = \frac{L + 2\bar{d}}{L + 3\bar{d}}  \frac{\bar{d}^{2}}{vL}
  \Big( e^{\frac{L}{\bar{d}}} - 1 -  \frac{L}{\bar{d}} \Big)
\end{displaymath}

\begin{displaymath}
    \tau_{p} = \frac{L}{v} + \frac{2\bar{d}}{v}
\end{displaymath}

where $ \bar{d} $ is the mean distance covered by a filament before
linking a motor; it can be related to the surface density of the
motors and other physical parameters.  $ v $ is the average speed of
filaments, and L their length. This gives the expression for $ r $

\begin{displaymath}
    r = 1 - \frac{\frac{L^2}{\bar{d}^2}\Big( 1 +
    3\frac{\bar{d}}{L}\Big)}{\Big( e^{\frac{L}{\bar{d}}} - 1 -
    \frac{L}{\bar{d}} \Big)}
\end{displaymath}

We have already said (page \pageref{lunghezza}) that microtubule
length enters through the parameters of our model.  The above formula
is a way to relate the filament length to the parameter $
r $ via independent, more detailed modeling.

Equations (\ref{eq:3}) \ldots (\ref{eq:6}) have the structure of
reaction-diffusion equations, with no source for $ H $, as filament
number is fixed.  The sources for $ M $, $ g $ and $ f $ arise
from the interaction between filament directions.

The existence of a source makes them different from other equations
obtained in the same way for two-species driven diffusive systems
(\cite{foster}, \cite{hwang}).  Another difference is that in this
last  case, having just two species and a field forcing in one spatial
direction, mean field equations come out to be one  dimensional, while
ours are not. However, by the general consideration  that patterns are
usually lower-dimensional (see \cite{halperin}), we expect that our
solutions will depend on one spatial variable, as well. In what
follows we will give more specific justifications of this fact.

\section{Steady states and stability}\label{sec.4}

A quick glance at our equations is sufficient to conclude that they
always admit the homogeneous solution
\begin{displaymath}
    \begin{array}{c} H = \bar{H}  \\ f,g = 0  \\ M = \bar{M} \:
        \textrm{which satisfies the Ising-like  equation} \\ \bar{M} +
        \mathrm{Th}(4J\bar{M}) = 0 \end{array}
\end{displaymath}
where, by definition of the parameters, $\vert \bar{M} \vert \leq
\bar{H}$.

This solution is the Gibbs equilibrium one when $ rE = 0 $.

It is interesting to notice that, when $ E = 0 $, even if $ r \ne 0
$, our equations can be derived from an equilibrium free energy; that
is, they have the form $ \mathbf{\dot{X}} = \frac{\delta F}{\delta
\mathbf{X}} $, with $ F $ a suitable functional of the fields.  In
particular, equations (\ref{eq:3}) and (\ref{eq:4}) are decoupled from
equations (\ref{eq:5}) and (\ref{eq:6}).  The first two describe a
relaxation process to the equilibrium phases of our model, while the
others admit as a unique stationary state the null field $ \vec{D} =
0  $.  These considerations cease to be true as soon as $ E \ne 0 $.

In fact, when $ rE \ne 0 $, nonzero irreversible currents  for $ f
$ and $ g $ arise:

\begin{equation}
    J_{g} = rE \bar{H} (1-\bar{H}+\bar{M}) \label{eq:7}
\end{equation}
and similarly for $ J_{f} $.  These stationary currents are the
mirror of a nonzero entropy production which we can measure
numerically (see sec.\ref{sec.5}), and, as with this last quantity,
are  linear in the nonequilibrium drive $ rE $.

\begin{figure}[!hbp]
    \centering
 \includegraphics{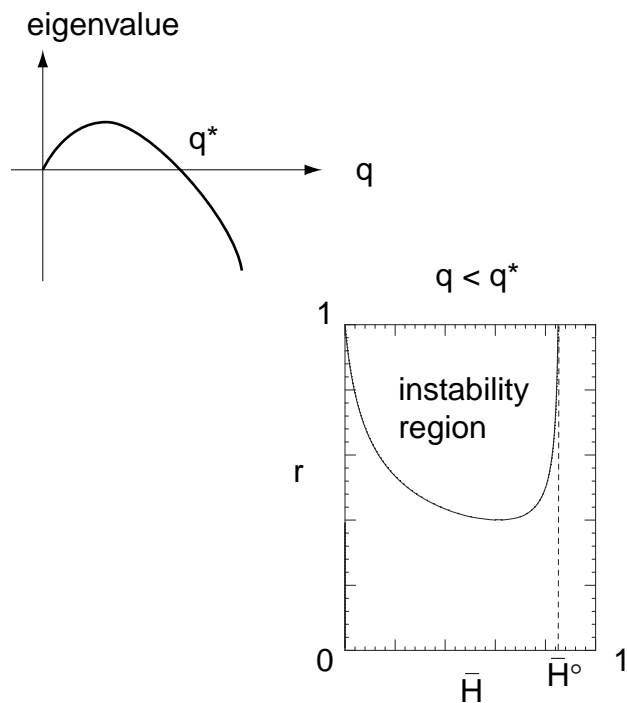} 
\caption{Linear
    stability analysis results. Our methods coincide  with those in
    \cite{halperin}. Current terms are kept to lowest  order. 
    \protect{ $ q $}
    is the wavelength of the perturbation. For 
    \protect{ $ q > q^{*}  $} the
    solution is always stable. For 
    \protect{ $ q < q^{*} $} we find an
    instability if 
    \protect{ $ \frac{\bar{H} (1-2\bar{H}+\bar{M})}{(1 -
    \bar{H}) (1 - \frac{\bar{M}}{2}) } > \frac{1 - r}{r^{2} E^{2}}$}.
    If \protect{ $ E $} and \protect{ $ M $} are fixed, the stability region looks like
    the one represented in the sketch above.}  \label{fig:3}
\end{figure}

We perform linear stability analysis around this homogeneous solution
(resumed in figure \ref{fig:3}), to find out that the system, far from
equilibrium, for perturbations of short wavelength, becomes unstable
in a region of the $ \bar{H},r $ plane. As can be seen from the
picture, for low concentration of filaments and low $ r $ the system
is always in a stable homogeneous state.  One instability may arise
after $ r $ reaches a critical value $ r_{c} $.

The next thing to do is to verify that this instability leads to the
inhomogeneous solutions we expect.  In doing this, we employ the same
techniques used in \cite{foster} and \cite{hwang}, that is, successive
substitutions leading to a nonlinear dynamical system for the variable
$ H $.  Let's outline our procedure.  By keeping terms to lowest
order in the fields one has a hint at what the solutions look like.
With this spirit, from equation (\ref{eq:3}) one obtains the relation

\begin{equation}
    \nabla H = rE H \vec{D} + \ldots \label{eq:8}
\end{equation}

When $ r \in (0,1) $ and the two dynamics are  actually competing,
Using (\ref{eq:8}) in eqn. (\ref{eq:4}) we can write an expansion of
$ M(x) $ around the Ising value $ \bar{M} $

\begin{equation}
    M = \bar{M} + \alpha [(1-\bar{M}) \partial_{x}^{2}H - (1+\bar{M})
    \partial_{y}^{2}H] +\alpha^2 \ldots \label{eq:9}
\end{equation}
with
\begin{displaymath}
    \alpha = \frac{1}{(1-r)(1 + \frac{J}{\cosh^{2}(J\bar{M})})}
\end{displaymath}

The last step is to use these two results in equations (\ref{eq:5})
and (\ref{eq:6}).

It is easy to acknowledge that the source terms  for $ g $ and $ f
$, are equal to zero iff $ g $ and $ f $  themselves are.  Thus,
the two fields are tendentially brought to zero  by the dynamics
(equations (\ref{eq:5}) and (\ref{eq:6})).

However, the decay times of $ f $ and $ g $ are very different.
Precisely, they are determined by the sign of $ M $.  What happens,
then, is that one of the two, say $ f $, goes rapidly to zero while
the other doesn't, so equation (\ref{eq:6}) allows studying the
stationary  states that depend on just one spatial variable, and we
can  substitute relations (\ref{eq:8}) and (\ref{eq:9}) in equation
(\ref{eq:5}) and obtain (similarly as in \cite{foster} and
\cite{hwang}) the ordinary differential equation for $ H $

\begin{equation}
   \Big(\frac{\partial H}{\partial x}\Big)^{2} = A H^{4} - B H^{3} + 2
    E^{2}  H^{2} - r^{2} E^{2} (1+\bar{M}) H - (1-r) H \log \vert H
    \vert \label{eq:10}
\end{equation}

where $ A $ and $ B $ are constants of integration. The solutions
of equation (\ref{eq:10}) belong to the class of cnoidal functions,
and are investigated in the mathematical literature. These are
periodic functions whose period and amplitude are related to each
other, by some known expressions. If the system has a finite size,
this constraint, toghether with minimization of energy, fixes the
vavelength of an inhomogeneous state.   

\begin{figure}[htbp]
    \centering \includegraphics{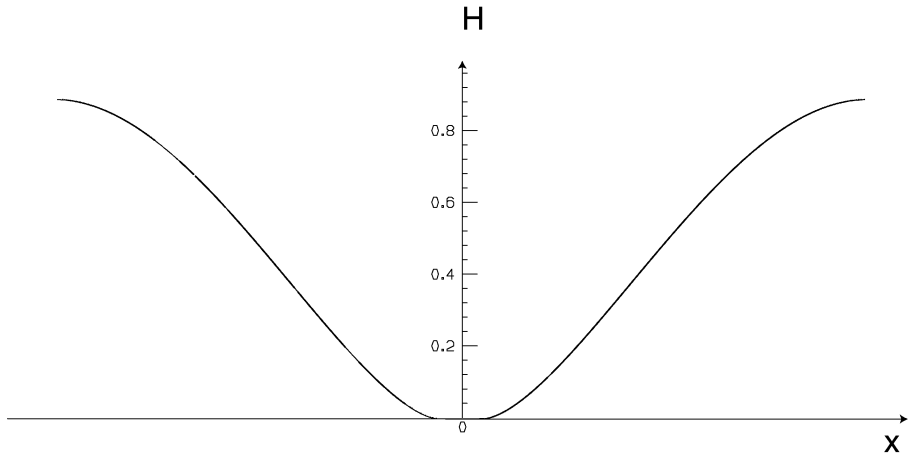} \caption{Plot of a
    solution of equation (\ref{eq:10}) obtained  with the boundary
    condition 
    \protect{ $ H \longrightarrow 1 $} as  
    \protect{ $ x \to  \pm \infty $}, which,
    given the total density, fixes the width  of the stripe.}
    \label{fig:grafico}
\end{figure}

\begin{figure}[btp]
    \centering \includegraphics{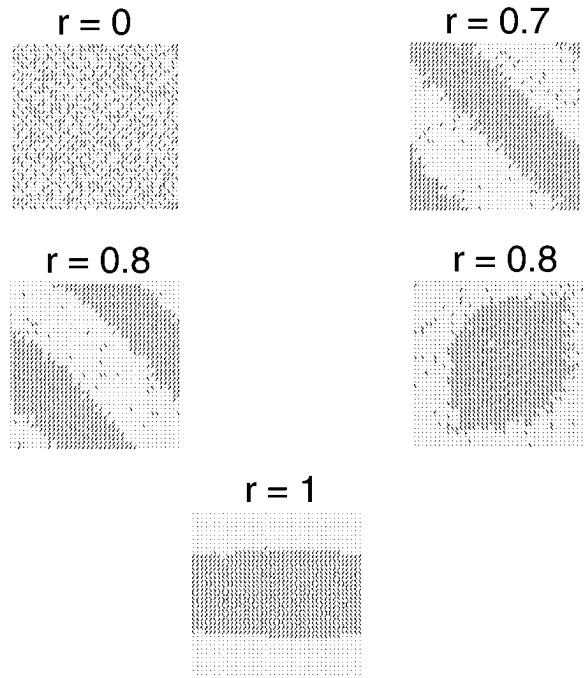} \caption{Steady
    states obtained with our simulations for 
    \protect{ $ J = 1  $}, 
    \protect{ $ K = 1.5$}, 
    \protect{ $ E = 8 $} and different values of 
    \protect{ $ r $}.  The case 
    \protect{ $ r =0 $} is the equilibrium state, 
    and the system is not sensitive to
    the value of \protect{ $E$}. For 
    \protect{ $ r = 0.7 $}  a stripe-like pattern is
    clearly identifiable. As \protect{ $ r $} becomes  closer to 
    \protect{ $ 1 $},
    droplet-like metastable states become more and  more long
    lasting. For 
    \protect{ $ r = 1 $} we obtain blocked states. These  results
    do not change sensibly as long as 
    \protect{ $ J > 0 $} and 
    \protect{ $  J,K $} are
    significantly lower (about one order of magnitude)  than 
    \protect{ $ E $}. \protect{ $ K $} can be positive or negative. } 
    \label{fig:st}
\end{figure}

Turning our attention to unbounded systems, we discover stripe-like 
states like that of figure \ref{fig:grafico}.
Our solutions are characterized by the fact that $ H $ may reach the
null value.  So, if $ r $ is small enough, independently from the
values of the coupling constants, there exist lines along which the
density reaches its maximum, compatibly with excluded volume.  Along
the same lines we find a discontinuity jump of the orientation field
$ \vec{D} $, whereas $ M $ fluctuates around the Ising value $
\bar{M} $ modulated by the second derivatives of $ H $ (Notice
that, at this stage, the constraint that $ M $ must always be lower
than $ H $ has to be implemented; we verify that it is satisfied for
suitable values of $ J $).  This behavior is confirmed by the
results of our numerical simulations.  For comparison, in figure
\ref{fig:st} are shown some snapshots of stationary states taken from
our simulations.
 
Finally, we want to note that, for $ r = 1 $ (and $ E \ne 0 $),
the four species are always simultaneously present and conserved, so
the expansion (\ref{eq:9}) around the Ising value of $ M $ ceases to
be significant, and is substituted by the relation

\begin{displaymath}
	\nabla \frac{M}{H} = E \hat{T} \nabla \frac{1}{H} + \ldots
\end{displaymath}

In this formula the presence of the conjugate operator $ \hat{T} $
imposes that the solutions depend on just one spatial variable, in
agreement with a pure multi-species driven diffusive system, and we
find so called ''blocked'' states present in the literature.

\section{Simulations. Entropy production}\label{sec.5}

We perform simulations at fixed total density equal to $ 1/2 $ on a
$ 40 \times 40 $ square lattice.  Montecarlo time unit $ \tau $
corresponds to 1600 moves, or one lattice sweep.  $ E $ is typically
one order of magnitude greater than $ J $ and $ K $.

\begin{figure}[htb]
    \centering \includegraphics{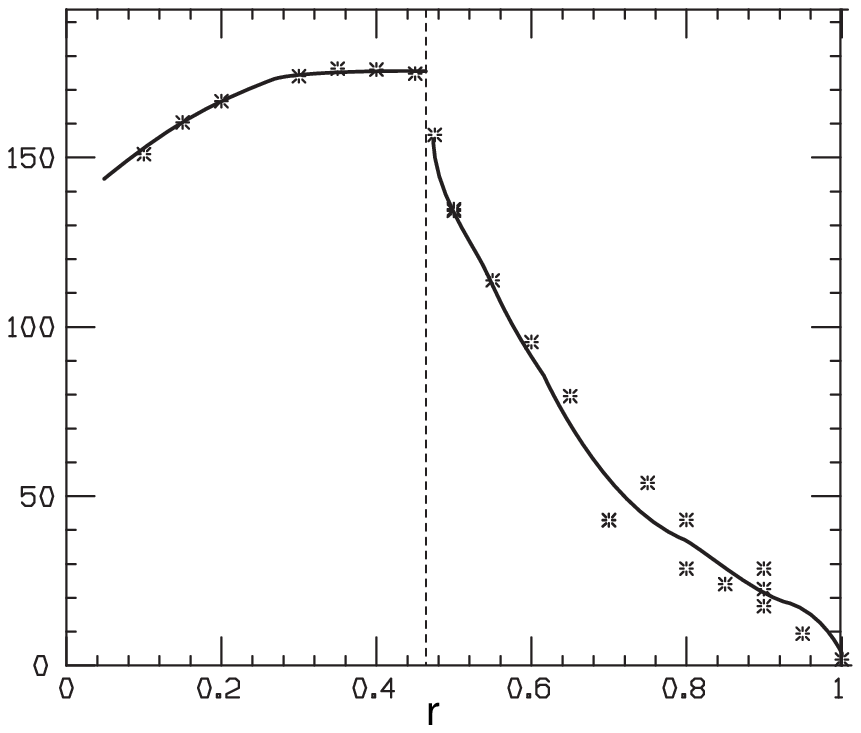} \caption{Accepted
    translations per sweep at the steady  state, mean value. The
    parameter is plot versus \protect{ $ r $}, and 
    \protect{ $ E  $} is fixed (
    \protect{ $J =1$}, \protect{ $ K = 1.5$}, \protect{ $ E = 8 $}).  
    The slope of the curve
    represents the mobility,  which, after reaching the saturating
    value of zero has a jump to  negative values, meaning that an
    increasing drive inhibits  translations.  The jump in this
    parameter corresponds to the  critical value for 
    \protect{ $ r $}. Doubling
    \protect{ $ E $} determines a change in the calculated points 
    of less than four percent (this is also valid for figures \ref{fig:5},
    \ref{fig:6} and \ref{fig:7}).}  
\label{fig:4}
\end{figure}

\begin{figure}[!htb]
    \centering \includegraphics{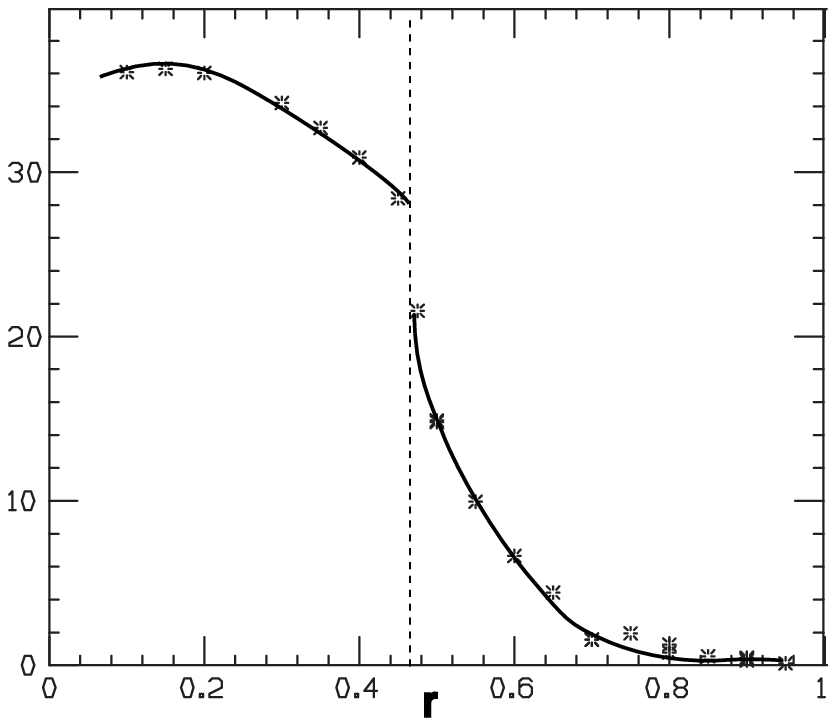} \caption{Mean value of
    accepted rotations per sweep versus \protect $ r $  at the steady state
    (\protect{ $J = 1$}, \protect{ $ K = 1.5$}, \protect{ $ E = 8 $}).   
    A decided drop of this
    parameter is observable at critical \protect{ $ r $}.}  \label{fig:5}
\end{figure}

Among the parameters we examine are the mean values of $ M $ and  $
\vec{D} $, and the mean number of accepted rotational and
translational Montecarlo moves (per sweep).  These latter quantities
measure the relative mobility of the system, that is, how much a
state is ``blocked'' (see figures \ref{fig:4} and \ref{fig:5}).

In order to calculate these means, data are sampled each $ 10^{3}
\tau $.  Total running time is about $ 10^{6} \tau $.  The time to
reach a steady state goes from $ 10^{4} $ to $ 10^{5} \tau $.
Inhomogeneous steady states typically arise for big enough values of
$ rE $ and have the form of a stripe either orthogonal or oblique
with respect to the direction chosen by the filaments.  An alternative
state is shaped like a droplet (of arrows with the same direction).
This last state is metastable for intermediate values of $ r $, but
its stability time diverges as $ r \to 1 $ (see figure
\ref{fig:st}).

To analyze these states, we pay particular attention to the entropy
production and the structure factor.

Following Lebowitz and Spohn (\cite{lebo}), we define the entropy
production at the stationary state as

\begin{displaymath}
    \frac{d}{dt} S_{irr} = \frac{d}{dt} S_{flux} = \frac{1}{t}
    \langle W(t) \rangle
\end{displaymath}

with

\begin{displaymath}
    W(t) = E \int_{0}^{t}\sum_{i = 1}^{4} \sum_{bonds}  J_{bond,i}(s)
    ds
\end{displaymath}

where $ J_{bond,i}$ is the net current of the $ i $-th species
along  the bond in the driven dynamics, and the above integral is just
a sum  over Montecarlo times.

\begin{figure}[!htbp]
    \centering \includegraphics{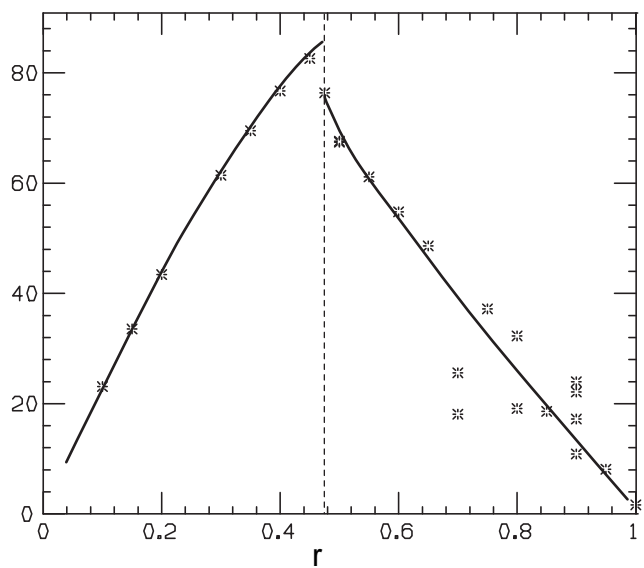} \caption{Entropy
    production at the steady state as a function of  
    \protect{ $ r $} (
    \protect{ $J = 1$}, \protect{ $ K = 1.5$}, \protect{ $ E = 8 $}).  
    After the parameter attains its
    maximum, there is a drop  and the system begins to structure. In
    the high \protect{ $ r $} region  the situation is more confusing because
    of the appearance of  metastable states. In fact droplets usually
    produce less entropy  than stripes. When \protect{ $ r = 1 $}, entropy
    production of the  steady states appears to be practically zero,
    even though the  states are not absorbing, because the microscopic
    dynamics is, in  principle, active. This effect is a consequence
    of the fact that  the driving field \protect{ $ E $} is very strong.}
    \label{fig:6}
\end{figure}

With this definition, we discover two distinct regimes (fig.
\ref{fig:6}).

In the first one, characterized by low $ r $, a class of homogeneous
entropy producing steady states is observable, and entropy production
increases almost linearly with $ r $ (see sec. \ref{sec.4}).

In the second regime inhomogeneous steady states appear, and entropy
production drops.  The critical value of $ r $ that separates these
two behaviors is also evident from the graphs of mean accepted moves
per sweep (figures \ref{fig:4} and \ref{fig:5}).  We estimate this
value to be $ r_{c} \simeq 0.45 $.  Considering the discontinuities
in the parameters, we believe that the transition is of the first
kind.

On the other hand, the structure factor is defined as

\begin{displaymath}
    S(k) = \langle \vert G(k) \vert^{2} \rangle
\end{displaymath}

with
\begin{displaymath}
    G(k) = \frac{1}{L} \sum_{x} \sigma^{4}_{x} \: e^{2 \pi \textrm{i}
    k  \cdot x}
\end{displaymath}

In particular, we observe the values assumed by $ S(k) $ in
correspondence with wave vectors $ k $ that are sensitive to
stripe-like states in various directions; namely $ k =
(0,1),(1,0),\frac{\sqrt{2}}{2}(1,1),\frac{\sqrt{2}}{2}(1,-1) $.  If
suitably normalized, these quantities take the value of unity when the
stationary state is a stripe in a well-defined direction; typical
values for droplet-like states are around $ 1/2 $.

Notice that time averaging of $ \vert G(k) \vert^{2} $ may be
meaningless if  the system explores a number of different
inhomogeneous steady states,  so one has to be very careful about the
state being stable.

\begin{figure}[!hbt]
    \centering \includegraphics{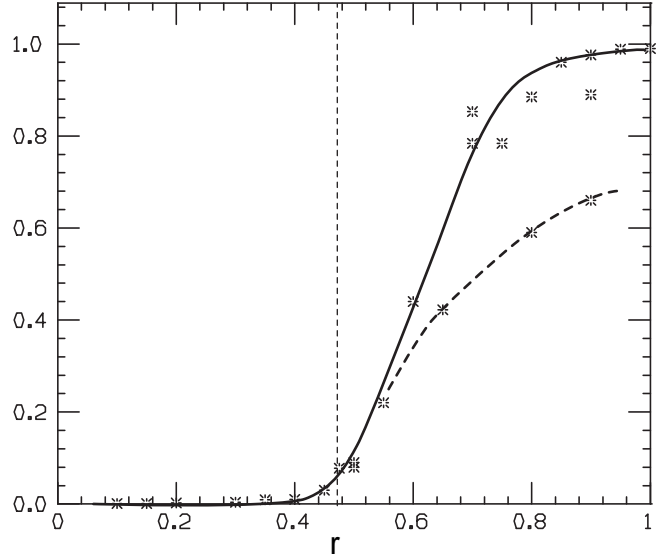} \caption{Structure
    factor as a function of \protect{ $ r $}  (
    \protect{ $J = 1$}, \protect{ $ K = 1.5$}, \protect{ $ E =
    8 $}) . This quantity  starts to rise when \protect{ $ r $} reaches its
    critical value. The dashed  line represents metastable
    droplet-like states, which become long  lasting as \protect{ $ r $}
    increases, and stable for \protect{ $ r = 1 $}. The  solid line stands for
    stripe-like states, independently on their  orientations. }
    \label{fig:7}
\end{figure}

Figure \ref{fig:7} is a plot of the structure factor of the steady
state as a function of $ r $.  In the high $ r $ region,
metastable states are subject to a slowing down of the dynamics and
become long lasting.  This is reflected by the two distinct behaviors
of the structure factor that can be seen in the picture (see also
figure \ref{fig:6}).

To sum up, our numerical studies confirm the results of mean field
analysis, and reveal a more intricate phenomenology for higher values
of $ r $. Due to the appearance of a slowing down in the dynamics,
we suspect the existence of a second phase transition in this region,
but more work is needed. In particular, a finite size scaling analysis
could be important.

\section{Links to self-organization assays}\label{sec.6}

We proceed by suggesting a possible interpretation of the  microscopic
model in terms of self-organization assays (mainly ref. \cite{ned1}).

In doing this there are a few main points to address.
First of all, our spins are driven one at a time whereas filaments
need to be at least two to be driven by the crosslinking soluble motor
complexes of \cite{ned1}. Second, motor complexes do not have a
fixed position, but can diffuse. Third, as the crosslinks between
filaments can produce a torque, a new question arises: do  we need
to take into account a driven rotational dynamics ? 

Dealing with the first two points is quite a simple matter. 

\begin{figure}[htbp]
    \centering
    \includegraphics{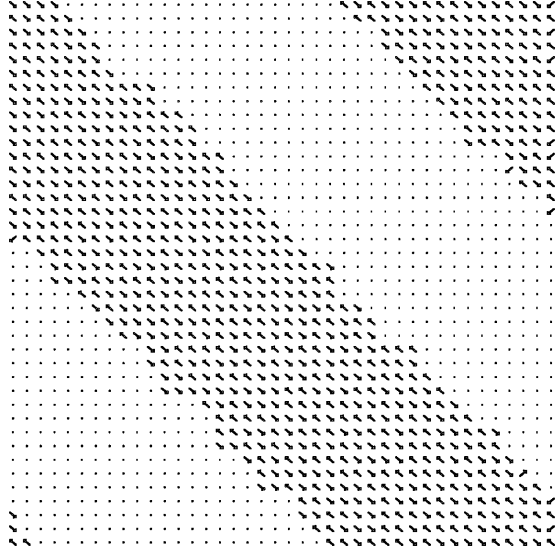}
    \caption{Stripe-like stationary state for the dynamics mimicking 
    self-organization assays. The values of the parameters are 
    \protect{ $ J = 1 $}, \protect{ $ K = -5 $}, \protect{ $ E = 2 $}, 
    \protect{ $ r = 1/2 $}.}
    \label{fig:dopmos}
\end{figure}

In particular, the first affects the elementary move of the
motors.  We have tested a different dynamics for motor drive
(\cite{marco}) which is more intuitively connectible with the
experiments in question \label{varia2}.  Referring to figure
\ref{fig:2}, the move of a motor is favoured by the field $ E(C) $
if the sites $ A $ and/or $ B $ are filled, and is an exchange
with the next nearest neighbor $ Z $.  Numerical results
show the emergence of inhomogeneous states and the same qualitative
behavior as the ordinary model.  New kinds of self-organized
states, as the stripe parallel to filament direction of figure
\ref{fig:dopmos} appear. The continuum mean field theory becomes a bit more
complicated, due to the appearance of terms in the powers of $ (1 -
H) $ in our system of equations. These terms come from the need of a
filled neighbor for a translation. Nevertheless, the solutions that we find remain 
valid in the limit of high filament density. 

As to the second point, at the scale we consider, 
motors act as a random field and it is not relevant whether this is a result
of brownian motion or of a static configuration of the motors.  

The third question is more essential and involves the
role of rotational driven dynamics in the process of
self-organization. 

The fact that the forcing torque privilegiates the
rotational moves aligning the filaments does not affect the nature
of the stationary states. In fact, in our ordinary model, inhomogeneous
states arise thanks to the joint effect of the  driven translations
and the aligning due to the coupling $ J $. So the torsional drive
is not in competition with equilibrium thermodinamics and can be absorbed in the
coupling constants $ J $ and $ K $, provided the system exhibits
orientational order. Only an analysis of the times of relaxation could
distinguish between the two behaviors.   

Finally, the task of connecting the effective parameters of our model with the
microscopic ones can be achieved with microscopic modeling of the kind
of ref. \cite{leibler}, specifically oriented on self-organization
assays.

\section{Conclusion}\label{sec.7}

We have introduced a model of nonequilibrium statistical mechanics
whose most interesting feature is the fact that the generalized force
is a dynamic, configuration dependent field.  This feature has been
inspired by the action of molecular motors on cytoskeletal filaments,
so we have made quantitative effort throughout the paper to interpret
our model in terms of systems that involve these objects.

Through analytic (mean field) approach and simulations, we have found
evidence for a nonequilibrium phase transition to inhomogeneous
states.  We were able to see that these inhomogeneous steady states
are genuinely far from equilibrium, checking their microscopic
currents and entropy production.  This results may be seen as a
theoretical evidence that a local drive in competition with diffusion and
excluded volume effects are sufficient to reach self-organized states. 

Work in progress on some variations of the model described in 
this paper make us confident on the generality of this statement. 
We already mentioned briefly two of these variations 
(pagg. \pageref{varia1} and \pageref{varia2}). Other numerical experiments 
have been tried for quasi three dimensional geometries, that is, 
allowing filaments to cross each other when diffusing with translations, 
and different moves for motor action, all showing the emergence of 
inhomogeneous stripe-like states. 

More work is needed to understand fully the statistical mechanical
properties of our model. In particular, we are working on three issues.    

First of all we want to perform a more careful analysis of the high $
r $ region, aiming to understand the nature of the stripe-like and droplet-like
states and their relation with the finite size of the lattice.

Secondly, we would like to know if the role played by the field $ M $  is
the same when one has a less radical discretization of microtubule
directions, for example if $ \sigma \in S^1 $.

Lastly, the dynamical properties of the relaxation of this model to its
stationary states are fully undiscovered. In particular, an analysis
of this kind could be useful to understand the role played by
torsional drive in dynamics resmbling self-organization assays.

\end{document}